# Low dose and fast grating-based x-ray phase-contrast imaging using the integrating-bucket phase modulation technique

Faiz Wali,[a] Shenghao Wang,[b,*] Huajie Han,[c] Kun Gao,[a] Zhao Wu,[a] Peiping Zhu,[d] and Yangchao Tian[a,#]

[a] National Synchrotron Radiation Laboratory, University of Science and Technology of China, Hefei, China, 230026

[b] Key Laboratory of Materials for High Power Laser, Shanghai Institute of Optics and Fine Mechanics, Chinese Academy of Sciences, Shanghai, China, 201800

[c] School of Engineering Science, University of Science and Technology of China, Hefei, China, 230027

[d] Institute of High Energy Physics, Chinese Academy of Sciences, Beijing, China, 10049

**Abstract:** X-ray phase-contrast imaging has experienced rapid development over the last few decades, and in this technology, the phase modulation strategy of phase-stepping is used most widely to measure the sample's phase signal. However, because of its discontinuous nature, phase-stepping has the defects of worse mechanical stability and high exposure dose, which greatly hinder its wide application in dynamic phase measurement and potential clinical applications. In this manuscript, we demonstrate preliminary research on the use of integrating-bucket phase modulation method to retrieve the phase information in grating-based X-ray phase-contrast imaging. Experimental results showed that our proposed method can be well employed to extract the differential phase-contrast image, compared with the current mostly used phase-stepping strategy, advantage of integrating-bucket phase modulation technique is that fast measurement and low dose are promising.

**Corresponding author:**
**\*** Shenghao Wang, E-mail: wangshenghao@siom.ac.cn
**#** Yangchao Tian, E-mail: ychtian@ustc.edu.cn

# 1 Introduction

Conventional X-ray imaging has found wide applications in medical imaging, social security and many other industrial fields. However, because of utilizing attenuation signal, contrast produced by weakly absorbing samples is poor, and this is the main reason that ruled out X-ray computed tomography for the investigation of musculotendinous trauma, infections, neoplasia and many other medical diagnosis (1). To overcome this problem, X-ray phase-contrast imaging was proposed as it could provide greatly improved contrast over conventional absorption-based imaging in biological samples, polymers and fiber composites (2-4). So far various X-ray phase-contrast imaging techniques have been provided in recent decades, these methods can be categorized as: crystal interferometer (5-7), free-space propagation (8, 9), diffraction enhanced imaging (10, 11) and grating interferometer (12-14).

Among these X-ray phase-contrast techniques, the phase signal retrieval is mostly achieved by the phase-stepping (PS) phase modulation technique. In crystal interferometer, a wedged phase shifter is usually placed in one of the beam path behind the beam splitter crystal, and by discontinuously moving the phase shifter step by step, the external input phase is varied and the interference fringes formed by the two beams would move, which would be applied to measure the sample's phase signal (5, 6). In the technique of diffraction enhanced X-ray phase-contrast imaging, typically the analyzer crystal is rotated step by step in a fixed angular interval to obtain the rocking curve, based on which the differential phase signal of the sample would be retrieved (10, 11). While in grating interferometer, PS is usually carried out by capturing a set of images at different positions of the analyzer grating (G2) (12-16), when G2 is scanned along the transverse direction, the signal intensity in each pixel of the detector oscillates as a function of the grating position, and by Fourier analysis of the intensity curve, the conventional absorption image, the refraction signal and the dark-field image of the sample can be simultaneously retrieved (14). The phase modulation of PS has the advantage of being easily performed and can be used successfully to retrieve the phase signal of the inspected sample, however, because of its discontinuous nature, PS has the fetal weakness of worse mechanical stability and high exposure dose, which greatly hinder its wide application in fast phase measurement and future clinical applications, where time resolution and low radiation dose are highly demanded.

In this manuscript, the phase modulation of integrating-bucket (IB) technique, which has been used more widely than PS in high-speed phase measurement at the wave range of visible light (17, 18), will be introduced to the field of grating-based X-ray phase-contrast imaging, in the aim of increasing the data acquisition speed and reduce the exposure dose delivered to the biological sample.

The following sections are arranged like this, firstly the physical concept of PS and IB will be demonstrated, and then the experimental setup will be described, after that, we will show the primary experimental results of evaluating the feasibility of employing IB method in grating-based X-ray phase-contrast imaging, finally, the potential usage of IB in high-speed X-ray phase-contrast computed tomography and other techniques of X-ray phase-contrast imaging will be discussed.

## 2  The concept of PS and IB phase modulation

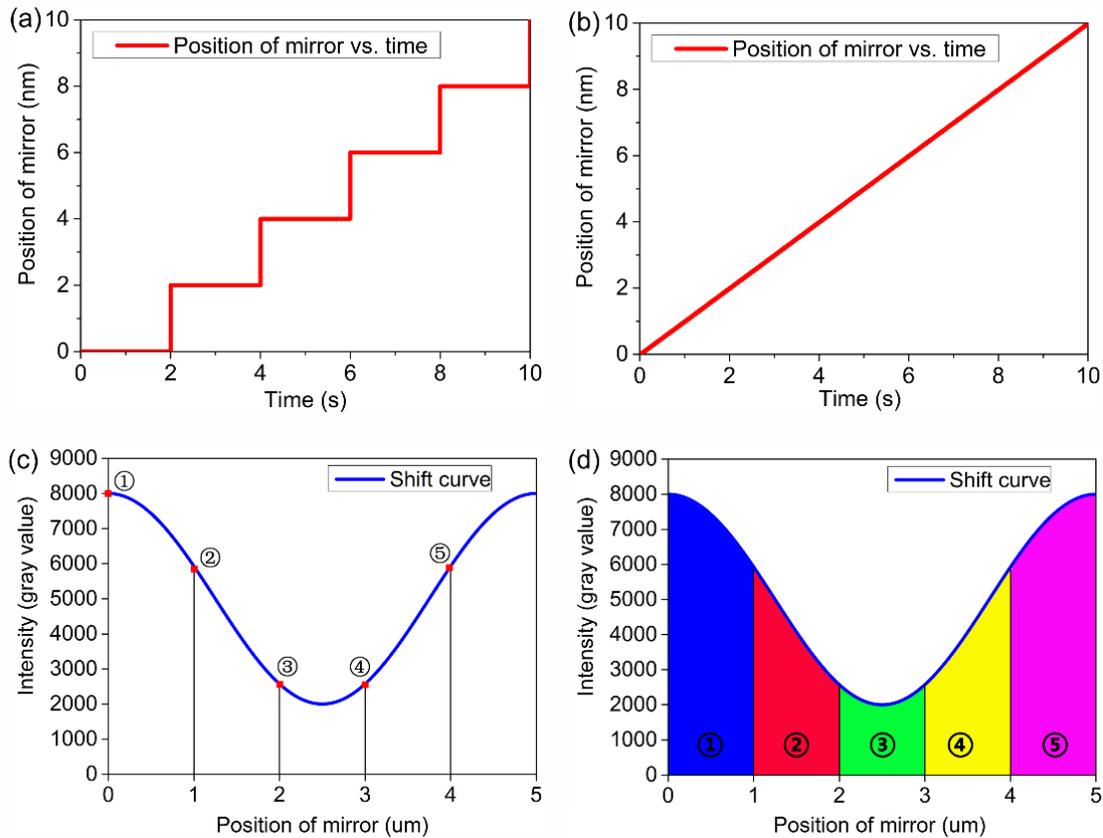

**Fig. 1** Demonstration of PS and IB phase modulations. (a) and (b) are respectively the curve of mirror's position vs. time in the scan of PS and IB, while (c) shows in PS the five positions of image acquisition along the shift curve, and in (d) the five color regions depict the five buckets of IB.

In the phase measurement technique at the wave range of visible light based on the classic Twyman-Green interferometer (18), for a certain pixel in the detector, with the moving of the PZT pushing mirror, its light intensity is changing as follow:

$$I_0 = I_0\left[1 + \gamma_0 \cos\left(\frac{x}{d} \times 2\pi + \phi\right)\right] \tag{1.1}$$

Where $I_0$ is dc intensity, $\gamma_0$ represent the modulation of the interference fringe, $x$ shows the position of the mirror, while $d$ is the period of the scan, and $\phi$ is the wave front phase at this pixel.

Detailed mathematical description of PS and IB phase modulations can be referred at (18), here we only use 5-step-PS and 5-bucket-IB for a simple demonstration of these two techniques. In the 5-step-PS, the mirror is moving discontinuously 5 steps as depicted in Fig. 1(a) the staircase curve of mirror's position vs. time, and the detector obtain 5 images, those are ①②③④⑤, at each step when the mirror is stopped and stabilized as shown in Fig.1(c), it is clear from equation (1.1) that the light intensity of the pixel at each step can be written as:

$$\left\{\begin{array}{l} I_1 = I_0\left[1 + \gamma_0 \cos(\phi)\right] \\ I_2 = I_0\left[1 + \gamma_0 \cos\left(\phi + \frac{2\pi}{5}\right)\right] \\ I_3 = I_0\left[1 + \gamma_0 \cos\left(\phi + \frac{4\pi}{5}\right)\right] \\ I_4 = I_0\left[1 + \gamma_0 \cos\left(\phi + \frac{6\pi}{5}\right)\right] \\ I_5 = I_0\left[1 + \gamma_0 \cos\left(\phi + \frac{8\pi}{5}\right)\right] \end{array}\right. \tag{1.2}$$

And the wave front phase $\phi$ at this pixel can be retrieved by the following equation:

$$\phi = \arg\left\{\sum_{k=1}^{5}\left[I_k \times \exp\left(2\pi i \times \frac{k}{5}\right)\right]\right\} \tag{1.3}$$

Where $I_k$ shows the light intensity at the $k^{th}$ step of the 5-step-PS scan.

In the 5-bucket-IB scan, the PZT pushing mirror is kept moving continuously in one period as depicted in Fig. 1(b) the line of mirror's position vs. time, and at the same time, the detector is also capturing photons continuously, five images as shown in Fig. 1(d) the five color regions, those are ①②③④⑤ are obtained with a certain integrating time, and the total times of the mirror's scan and the detector's data acquisition are equal, we can see from equation (1.1) that the intensity of the pixel at each image can be written as:

$$\begin{cases} I_1 = \frac{5}{d}\int_0^{d/5} I_0 \left\{1 + \gamma_0 \cos\left[\phi + \frac{x}{d} \times 2\pi\right]\right\} dx \\ I_2 = \frac{5}{d}\int_{d/5}^{2d/5} I_0 \left\{1 + \gamma_0 \cos\left[\phi + \frac{x}{d} \times 2\pi\right]\right\} dx \\ I_3 = \frac{5}{d}\int_{2d/5}^{3d/5} I_0 \left\{1 + \gamma_0 \cos\left[\phi + \frac{x}{d} \times 2\pi\right]\right\} dx \\ I_4 = \frac{5}{d}\int_{3d/5}^{4d/5} I_0 \left\{1 + \gamma_0 \cos\left[\phi + \frac{x}{d} \times 2\pi\right]\right\} dx \\ I_5 = \frac{5}{d}\int_{4d/5}^{d} I_0 \left\{1 + \gamma_0 \cos\left[\phi + \frac{x}{d} \times 2\pi\right]\right\} dx \end{cases} \quad (1.4)$$

Through a mathematical derivation, it would be founded that the wave front phase $\phi$ of this pixel can be retrieved by the same algorithm as written in equation (1.3).

## 3 Material and method

### 3.1 Experimental setup

The grating-based X-ray phase-contrast imaging experiments were carried out at the National Synchrotron Radiation Laboratory of the University of Science and Technology of China. Fig. 2(a) is the mechanical structure of the imaging setup, it mainly consists of an X-ray tube, an X-ray flat panel detector and three micro-structured gratings, which are built on multi-dimensional motors (Beijing Optical Century Instrument Co., Ltd, China), an ultra-precision piezoelectric translation stage (Micronix Inc., California, America.) with an encoder has been used for scanning G2. The X-ray tube is a cone beam source (YXLON international GmbH, Hamburg, Germany) with round focal spot (1.0 mm in diameter) on a tungsten target anode, its tube voltage can be adjusted from 7.5 kV to 160 kV, and the X-ray tube is cooled using a commercial available centrifugal chiller (HTCY Technology, Beijing, China). The source grating (G0) (period 100 μm, gold height 50 μm, size 1×1 cm²) is positioned about 10 mm from the emission point inside the X-ray source. The projection grating (G1) (period 50 μm, gold height 50 μm, size 10×10 cm$^2$) is placed 270 mm behind G0. G2 (period 100 μm, gold height 50 μm, size 10×10 cm$^2$) is positioned in contact with the flat-panel detector and the distance between G1 and G2 is 270 mm. All the three gratings were fabricated by the LIGA process, involving EUV photo etching and electroplating. The X-ray image was captured using a flat penal detector (PerkinElmer Inc., Waltham, Massachusetts, USA) with an effective receiving area of 20.48×20.48 cm$^2$ and 0.2×0.2 mm$^2$ pixel size. The control of the motorized stages, the piezoelectric stage and the flat panel detector are achieved with a custom developed LabVIEW software (19).

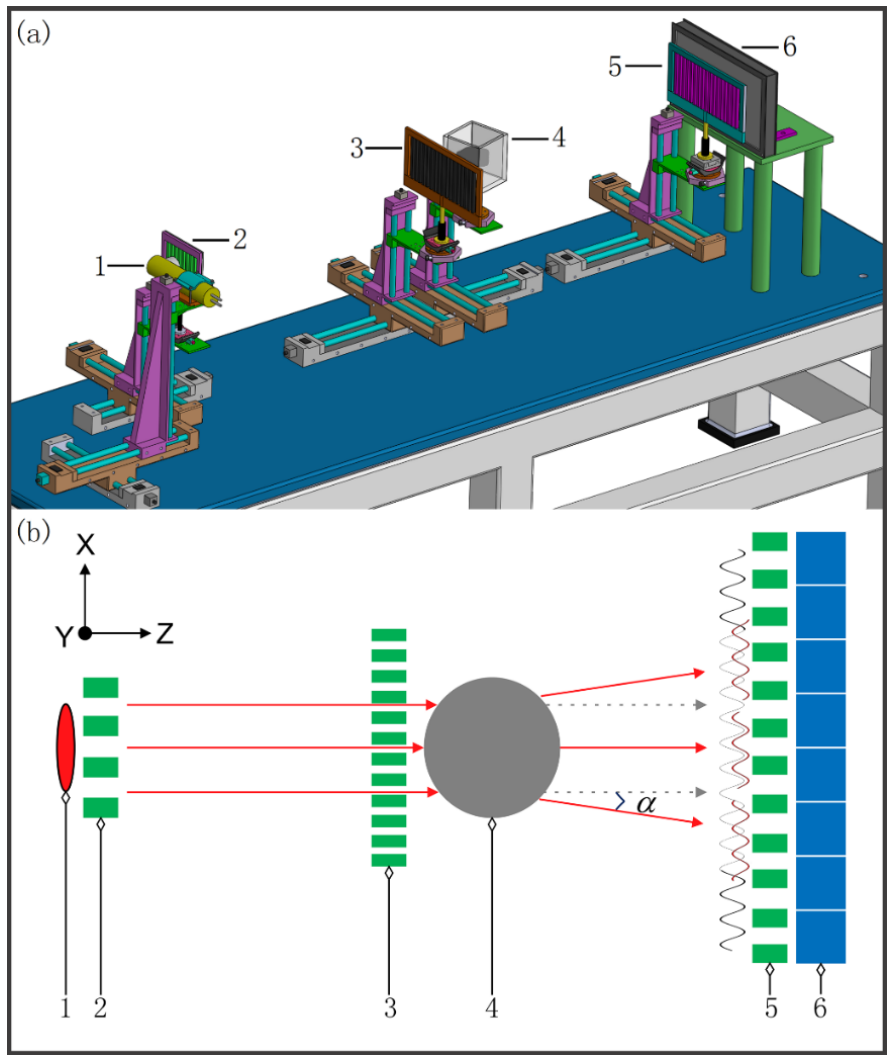

**Fig. 2** (Color online). Mechanical structure (a) and working principle (b) of the grating-based x-ray phase-contrast imaging setup. 1. X-ray source, 2. Source grating, 3. Projection grating, 4. Sample, 5. Analyzer grating, 6. X-ray flat panel detector.

## 3.2 Working principle

As illustrated in Fig. 1(b), the G0 grating, an absorbing mask with transmitting slits, set close to the X-ray tube anode, produces an array of line sources. G1 works as a projection grating, via direct geometric projection, shadow of G1 forms in the plane of G2, and finally moire fringe emerges at the receiving surface of the detector. The differential phase-contrast image information process achieved by the system essentially relies on the fact that the sample placed in the X-ray beam path induces a slight refraction of the beam transmitted through the object, and the fundamental idea of the differential phase-contrast imaging depends on locally detection these angular deviations. The angle $\alpha$ is proportional to the local gradient of the object's phase shift, and it can be written as:

$$\alpha = \frac{\lambda}{2\pi} \times \frac{\partial \Phi(x,y)}{\partial x} \qquad (1.5)$$

Where $\Phi(x,y)$ is the phase of the wave front and $\lambda$ stands for the wavelength of the radiation, determination of the refraction angle can be achieved traditionally by PS technique (19, 20).

## 4 Experimental results and data analysis

The sample we used contains two PMMA and two POM cylinders, each having a diameter of 5 mm and 10 mm, respectively. After fine alignment of the three gratings and the sample stage, the X-ray tube was operated at a voltage of 50 kV and current of 22.5 mA.

In the first set of experiment, the data was collecting by traditional PS method, in which 104 steps were adopted, and for each step, 30 raw images were captured to reduce statistical and systemic noise, noted that the exposure time of each raw image is 2 second. The PS scan was repeated twice before and after inserting the sample into the beam path, respectively.

The sample's refraction image generated by the PS method was computed by a LabVIEW software (19) with the following algorithm (21).

$$\alpha(m,n) = \frac{p_2}{2\pi d} \times \arg \left\{ \frac{\sum_{k=1}^{N} \left[ I_k^s(m,n) \times \exp\left(2\pi i \times \frac{k}{N}\right) \right]}{\sum_{k=1}^{N} \left[ I_k^b(m,n) \times \exp\left(2\pi i \times \frac{k}{N}\right) \right]} \right\} \qquad (1.6)$$

Where $N$ is the number of steps, $I_k^s(m,n)$ and $I_k^b(m,n)$ are the gray value of pixel $(m,n)$ at the $k^{th}$ step with and without sample, respectively, $p_2$ is the period of G2, and $d$ represents the distance between G1 and G2.

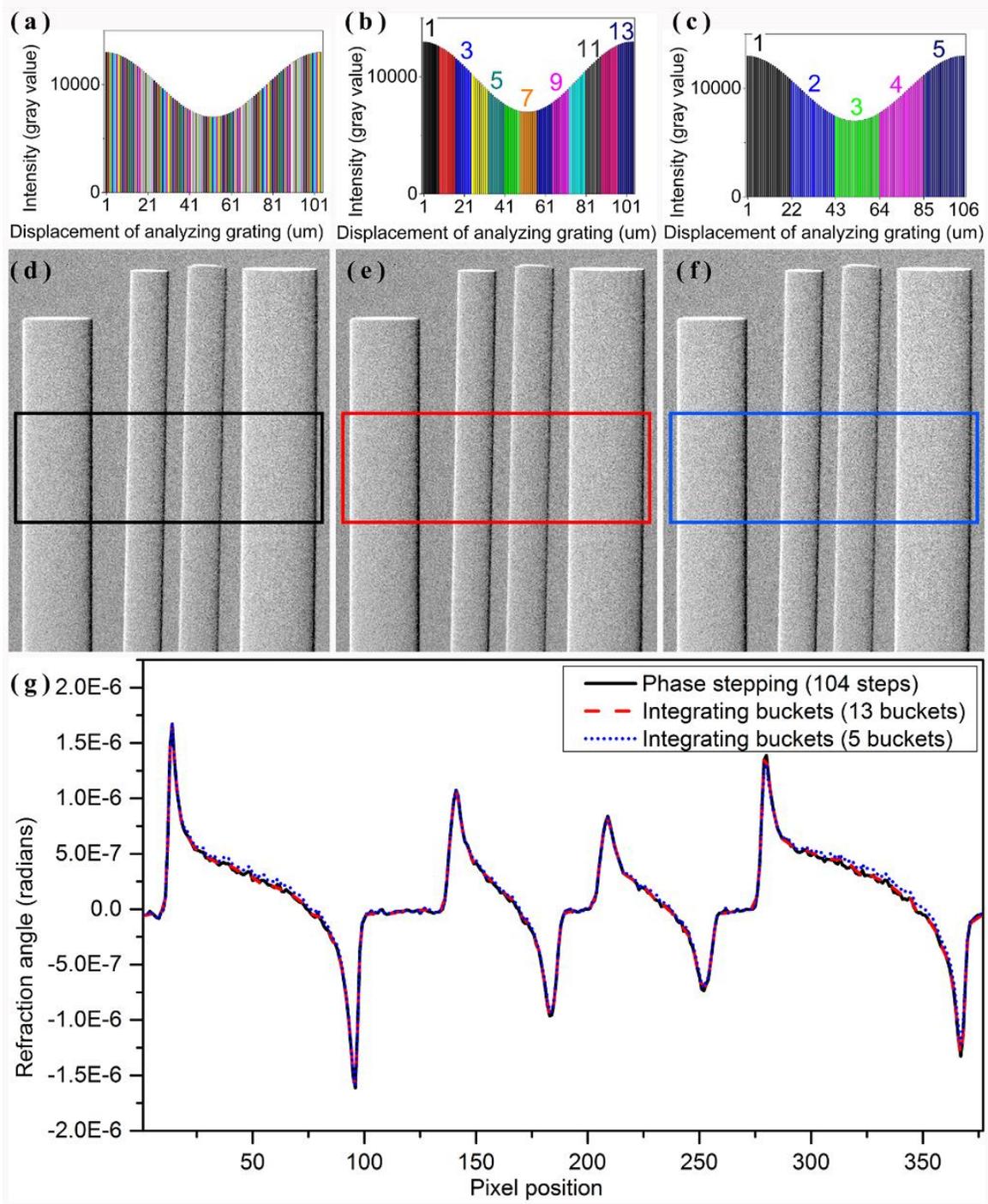

**Fig. 3** Result of the first set of experiment. (a) demonstrates the shift curve produced by the 104-step-PS scan, while (b) and (c) shows respectively the partition of the 104 steps into 13 and 5 buckets, (d) is the refraction

image computed by the 104-step-PS scan, (e) and (f) is the phase-contrast image generated respectively by IB method containing 13 and 5 buckets, in (g) the color lines shows the profile of refractive angle in the corresponding rectangles in (d), (e) and (f).

As demonstrated in Fig. 3(a) is the shift curve plotted by the 104 steps, and Fig. 3(d) is the refraction image computed by the 104-step-PS scan. Considering that in the 104-step-PS scan, the number of steps we chosed is remarkably large and the distribution of the data point on the shift curve is very dense as shown in Fig. 3(a), we think this set of raw data can be approximately handled using the IB method. Fig. 3(b) shows the partition of the 104 steps into 13 buckets, which was managed as follow, the images from $1^{st}$ to $8^{th}$ in the 104 steps were averaged as the $1^{st}$ bucket, and the images from $9^{th}$ to $16^{th}$ were averaged as the $2^{nd}$ bucket, the rest can be done in the same manner, and finally the images from $97^{th}$ to $104^{th}$ were averaged as the $13^{th}$ bucket. Fig. 3(e) is the differential phase-contrast image generated by IB method containing 13 buckets using equation (1.6). Similarly Fig. 3(c) shows the partition of the 104 steps into 5 buckets, and Fig. 3(e) is the refraction image produced by the IB method containing 5 buckets, it should be pointed out that the $5^{th}$ bucket was averaged by only 20 images (the former 4 buckets are generated by 21 images) from $85^{th}$ to $104^{th}$. In Fig. 3(g) the three color lines show respectively the profile of the refractive angle in the rectangles as depicted in Fig. 3(d), Fig. 3(e) and Fig. 3(f), noted that the 200 rows were averaged to compute the profile, in the aim of reduce the image noise.

In the second set of experiment, firstly, 10-step-PS scan was performed, and at each step, 49 raw images were averaged to reduce the image noise, the exposure time of each image is 2 second. The PS scan is repeated twice before and after inserting the sample into the beam path respectively, and then, the phase modulation of IB method containing 10 buckets was employed, for the $1^{st}$ bucket, the analyzer grating was moving back and forth 5 times with a constant speed from the initial position (x=0 um) to x=10 um, and at the same time, the detector began to capture photons and 49 raw images were obtained and averaged, the total exposure time is 98 seconds, which is the same as the time consumed to move the analyzer grating, after that, the same process was repeated from x=10 um to x=20 um, from x=20 um to x=30 um ···and finally from x=90 um to x=100 um, the phase-contrast image generated by the PS and IB methods are both computed by equation (1.6).

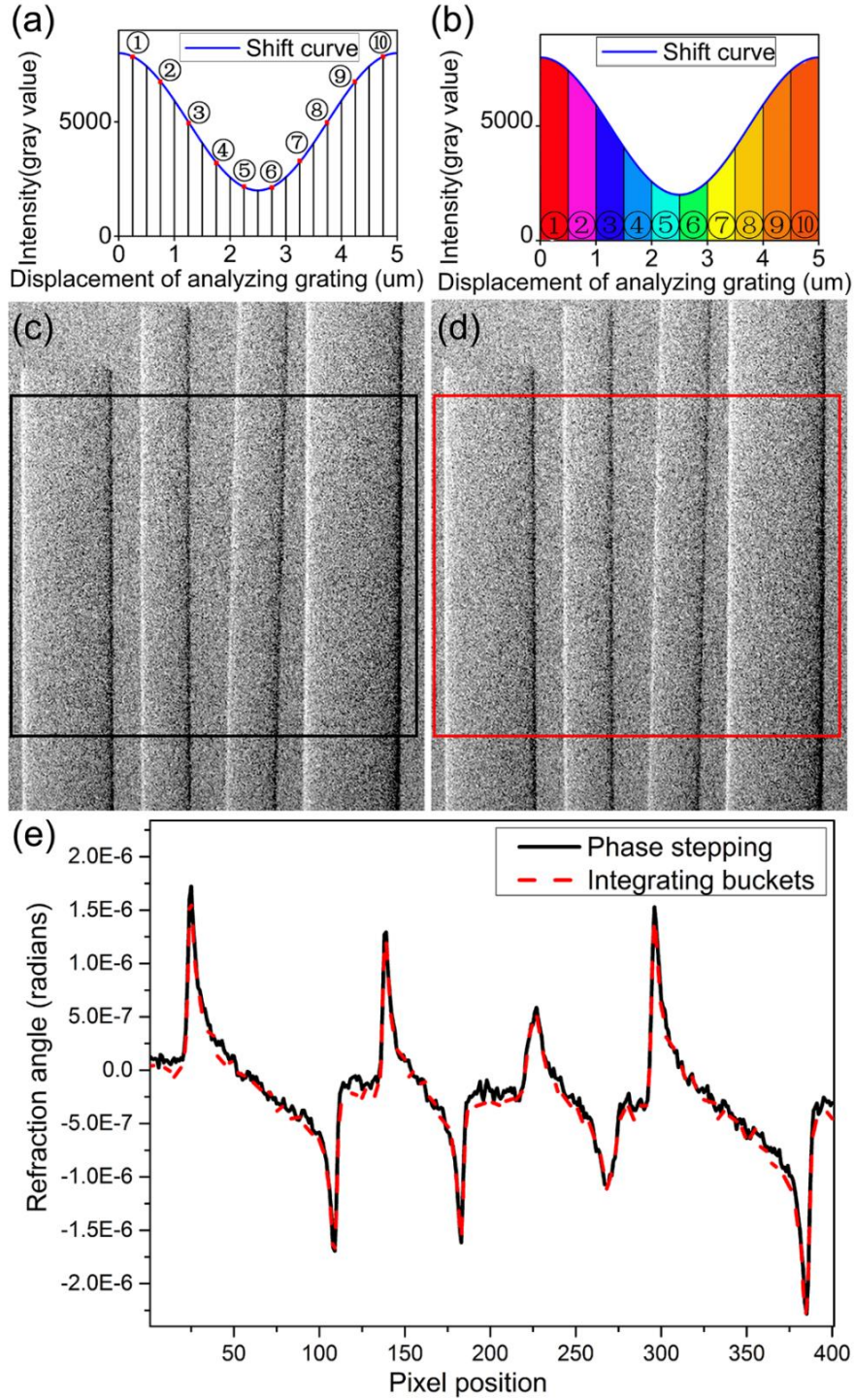

**Fig. 4** Result of the second set of experiment. (a) demonstrates the 10-step-PS scan, while (b) shows the technique of 10-bucket-IB, (c) and (d) depict the refraction image computed by PS and IB, respectively, in (e) the lines shows the profile of the refractive angle.

Fig. 4 shows the generated refraction images and the quantitative result. Fig. 4(a) and Fig. 4(b) demonstrate respectively the techniques of 10-step-PS and 10-bucket-IB, and as shown in Fig. 4(c) and Fig. 4(d) are the refraction image computed by the two methods, in Fig. 4(e) the color lines show respectively the profile of the refractive angle in the rectangles as depicted in Fig. 4(c) and Fig. 4(d), noted that the 200 rows were averaged to compute the profile.

By visually comparing the three refraction images as shown in Fig. 3(d), Fig. 3(e) and Fig. 3(f), which are respectively generated by 140-step-PS, 13-bucket-IB and 5-bucket-IB, and meanwhile the two refraction images as shown in Fig. 4(c) and Fig. 4(d) produced respectively by 10-step-PS and 10-bucket-IB scan, also the profiles of the refractive angle depicted in Fig. 3(g) and Fig. 4(e), it can be concluded that in contrast with the traditional phase modulation of PS, the proposed IB technique could deliver almost the same phase-contrast image.

## 5 Discussion

The aforementioned experiments show the primary usage of IB phase modulation method in grating-based X-ray phase-contrast imaging, and it should be point out that IB method could be well merged into the technique of X-ray phase tomography. In order to reduce the exposure dose delivered to the patient, high-speed X-ray phase tomography is very important in clinical diagnose, and it also has huge application prospect in time-resolved observation of the dynamic process of biological soft tissues, for example monitoring the respiratory process of the human lungs, by now many fast measuring techniques of X-ray phase tomography have been proposed: (a) The time-resolved X-ray phase-contrast computed tomography was realized by a single shot strategy and excellent experiments were reported (22, 23), however low spatial resolution is its fatal defect; (b) By continuously rotating the sample and employing the traditional PS scan technique, fast X-ray phase tomography is also achieved (24-26), but it is very clear that the discontinuity nature of PS greatly limits its data acquisition speed.

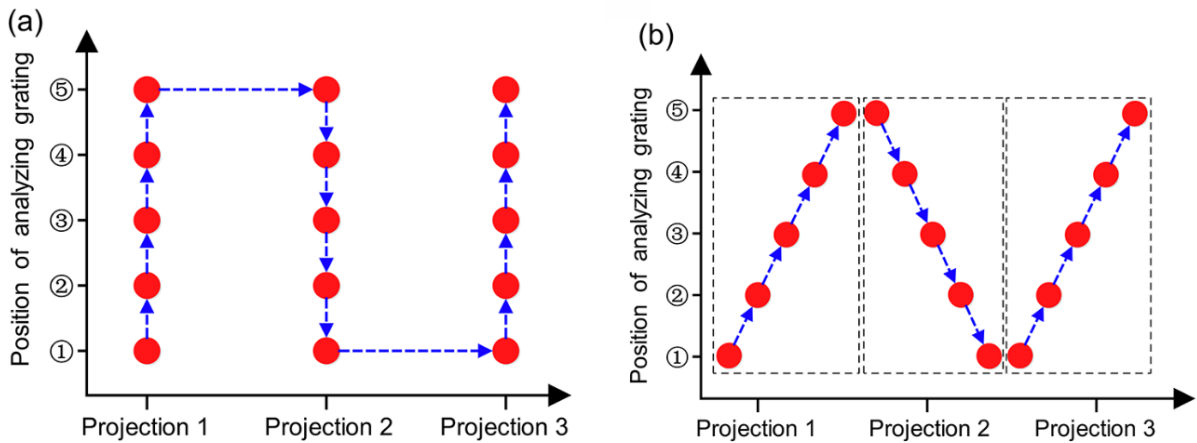

**Fig. 5**. Demonstration of the idea of combing IB with CT scan. (a) shows a fast manner, and (b) depicts a ultrafast strategy.

The combination of IB phase modulation technique with CT scan can be easily and perfectly performed in two ways as demonstrated in Fig. 5 to realized time-resolved x-ray phase-contrast computed tomography: (a) as shown in Fig. 5(a), the analyzer grating is moving back and forth continuously and the sample is rotating discontinuously step by step; (b) both the analyzer grating and the sample are moving continuously in a concerted speed as demonstrated in Fig. 5(b), we believe our proposed scheme would bring forth new and valuable technique of fast X-ray phase tomography.

Another idea of combining IB with CT scan we think the most ideal is introducing the sinusoidal IB phase modulation technique into the CT scan procedure, the concept of sinusoidal IB phase modulation has also been used in fast phase measurement at the wave range of visible light (27-30), and its key feature is that the motion of analyzer grating is harmonic vibration, which would perfectly meet the nature of reciprocating motion and the highest measuring speed is thus promising.

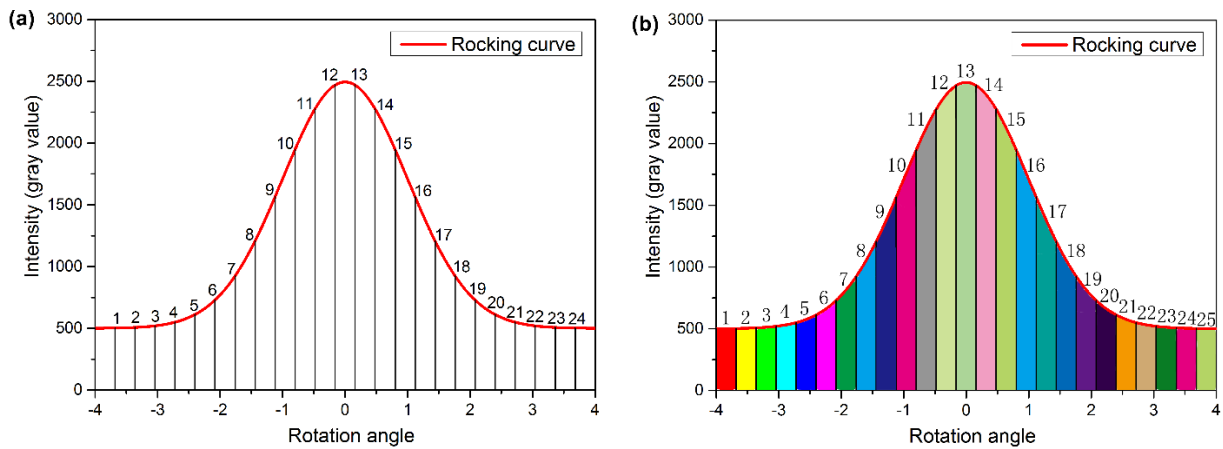

**Fig. 6** Conception of using IB technique in diffraction enhance X-ray phase-contrast imaging, (a) shows the usage of traditional PS method to obtain the rocking curve, while (b) demonstrates the idea of employing the IB method.

The method of IB can also be used in other X-ray phase-contrast imaging techniques, in crystal interferometer the implementation can be realized almost in the same manner with that in grating-based X-ray phase-contrast imaging, and in diffraction enhance X-ray phase-contrast imaging (10, 11), as shown in Fig. 6(a) is the usage of traditional PS method to obtain the rocking curve while rotating the analyzer crystal, and Fig. 6(b) demonstrates the idea of employing the IB method, it is very clear that the usage of IB will remarkably increase the data acquisition speed and the exposure dose delivered to the biological sample is therefore can be reduced, however it should be pointed out that unlike the sinusoid shift curve in grating-based and crystal interferometer based phase-contrast imaging, the rocking curve in diffraction enhance X-ray phase-contrast imaging is typically regarded as Gaussian curve, slight difference would exist in the retrieved phase signal produced by the traditional PS and the IB method.

6. **Conclusion**

In conclusion, the phase modulation of integrating-bucket technique was introduced to grating-based X-ray phase-contrast imaging in this manuscript, and our preliminary experimental results show its good feasibility. Compared with the current mostly used technique of phase-stepping, advantage of integrating-bucket phase modulation method is that low dose and fast X-ray imaging are promising.

**Acknowledgments**

This work was partly supported by National Research and Development Project for Key Scientific Instruments (Grant No. CZBZDYZ20140002), National Natural Science Foundation of China (Grant No. 61705246) and Fundamental Research Funds for the Central Universities (WK2310000065). The authors want to thank the Chinese Academy of Sciences and The World Academy of Sciences (CAS-TWAS) President's Fellowship program for financial support.